\documentstyle[12pt]{article}
\input epsf

\begin{document}
\baselineskip=15pt
\newcommand{\x}{{\bf x}}
\newcommand{\y}{{\bf y}}
\newcommand{\z}{{\bf z}}
\newcommand{\bp}{{\bf p}}
\newcommand{\A}{{\bf A}}
\newcommand{\B}{{\bf B}}
\newcommand{\p}{\varphi}
\newcommand{\del}{\nabla}
\newcommand{\be}{\begin{equation}}
\newcommand{\ee}{\end{equation}}
\newcommand{\bq}{\begin{eqnarray}}
\newcommand{\eq}{\end{eqnarray}}
\newcommand{\ba}{\begin{eqnarray}}
\newcommand{\ea}{\end{eqnarray}}
\def\r{\nonumber\cr}
\def\hf{\textstyle{1\over2}}
\def\qr{\textstyle{1\over4}}
\def\Sc{Schr\"odinger\,}
\def\sc{Schr\"odinger\,}
\def\'{^\prime}
\def\>{\rangle}
\def\<{\langle}
\def\-{\rightarrow}
\def\dbd{\partial\over\partial}
\def\tr{{\rm tr}}

\begin{titlepage}
\vskip1in
\begin{center}
{\large Order $1/N^2$ test of the Maldacena conjecture: Cancellation of the
  one-loop Weyl anomaly.} 
\end{center}
\vskip1in
\begin{center}
{\large
Paul Mansfield and David Nolland

Department of Mathematical Sciences

University of Durham

South Road

Durham, DH1 3LE, England}

{\it P.R.W.Mansfield@durham.ac.uk}, {\it D.J.Nolland@durham.ac.uk}

\end{center}
\vskip1in
\begin{abstract}

\noindent 
We test the Maldacena conjecture for type IIB String Theory/
${\cal N}=4$ Yang-Mills by
calculating the one-loop corrections in the bulk theory 
to the Weyl anomaly of the boundary CFT when the latter is coupled to a 
Ricci flat metric. The contributions cancel within each supermultiplet, in
agreement with the conjecture.
\end{abstract}

\end{titlepage}

The truly remarkable nature of the AdS/CFT correspondence is nicely
illustrated by Henningson and Skenderis' beautiful computation (from the
bulk) of the leading order contribution to the boundary theory Weyl anomaly
\cite{Henningson}. This test of
Maldacena's conjecture \cite{Maldacena} demonstrates that, 
at large $N$, an (exact) one-loop calculation in four-dimensional
${\cal N}=4$ Super-Yang-Mills theory
can be reproduced by a tree-level calculation in General Relativity.

When the boundary Yang-Mills theory is coupled to a non-dynamical, external
metric, $g_{ij}$, the Weyl anomaly, ${\cal A}$, is the response of the free
energy, $F[g_{ij}] $, 
to a scale transformation of that metric: 
\be\delta F=\int d^4x\,{\sqrt g}
\delta\rho {\cal A},\qquad 
\delta g_{ij}=\delta\rho g_{ij}.\ee

On general grounds ${\cal A}$ must be a
linear combination of the Euler density, $E$, and the square of the Weyl
tensor, $I$. A well-known one-loop calculation gives 
\be\label{cfta}\pi^2{\cal A}=-(N^2-1)(E+I),\ee
and supersymmetry protects this from higher-loop corrections.
The tree-level calculation in the bulk reproduces the leading $N^2$ piece,
by solving the Einstein equations perturbatively near the boundary.
We would expect that the $-1$ piece is due to loops in the bulk, but 
these depend on much more than just classical General Relativity, and
reproducing them provides a more stringent test of the Maldacena conjecture
sensitive to the detailed particle content of the bulk theory.

Other authors have discussed $O(N)$ contributions which arise in slightly
different theories \cite{blau,odinsov}, but in the current case they are not
present and we will be interested in the $O(1)$ contribution, which
corresponds to a one-loop effect in the bulk.

In this letter we describe the results of the one-loop calculation in the bulk 
when the boundary metric is Ricci flat. This restriction enables us to
avoid having to solve for the metric perturbatively; the bulk metric 
\be
ds^2={1\over t^2}\left( l^2\,dt^2+\sum_{i,j}g_{ij}\,dx^i\,dx^j\right),\qquad t>0
\label{newmet}
\ee
satisfies the Einstein equations with cosmological constant $\Lambda=-6/l^2$.
For a Ricci-flat metric $E=-I=R^{ijkl}R_{ijkl}/64 $, so that (\ref{cfta})
vanishes. Our computation gives the coefficient of the square of the Riemann
tensor, which is one half of the information needed to fix the Weyl
anomaly completely. For the
Maldacena conjecture to hold this must be zero. We work in the
field theory 
limit of String Theory, but we will argue that corrections from higher string
modes vanish.

\medskip
The spectrum of the compactification on $AdS_5\times S^5$ of 
Type IIB Supergravity was obtained in \cite{Kim} by perturbing 
classical equations of motion about a solution in which all the fields
vanish except the metric and a self-dual five-form field strength.
These classical equations do not follow from an action principle,
but rather are chosen to be compatible with supersymmetry.
(If there had been an action this calculation would have given its 
second functional derivative.) 

Since we work with a more general background
than the maximally symmetric $AdS_5$ we need to repeat the calculation.
We have done this, and also constructed a corresponding Lagrangian 
which we need in order to go beyond the classical level and quantize the
theory. 
The main complication is that some of the equations of motion used in
\cite{Kim} 
are no longer present when we impose all the gauge conditions, as we must in
order to do one-loop calculations.
We constructed our Lagrangian {\em after}
expanding in $S^5$ spherical harmonics, so it is 
local in $AdS_5$, but {\em not} in the full ten-dimensional space. In this
way we avoid
the paradox associated with a Lagrangian description of self-dual
field strengths. 

The conclusion of our analysis, perhaps
unsurprisingly, is that the spectrum of physical particles is the 
same as in \cite{Kim}, so we leave a detailed description of the calculation
to another place.

\medskip
 
The central object of interest in the AdS/CFT correspondence
is the `partition function' given as a functional integral for the bulk
theory in which the fields have precribed values, $\varphi$, on the
`boundary' at
$t=0$ \cite{Witten}. Because of the divergences in the metric it is necessary to introduce 
a small $t$ cut-off, $\tau'$, even in tree-level calculations. 
At one-loop it is also necessary to introduce a large $t$ cut-off $\tau$;
this introduces another boundary, and the functional integral should be 
performed with the fields taking prescribed values, $\tilde\varphi$, 
there as well. Consequently the partition function is the limit as 
the cut-offs are removed of a functional
$\Psi_{\tau,\tau'}[\tilde\varphi,\varphi]$. The exponential of the
free energy is the field independent part of this partition function. 
With the regulators in place the free energy becomes a function of 
$\tau,\tau'$, and the Weyl anomaly can be found by exploiting the
invariance of the five-dimensional metric (\ref{newmet}) under $t\rightarrow (1+
\delta\rho/2)t$, $g_{ij}\rightarrow (1+
\delta\rho)g_{ij}$, with $\delta\rho$ constant. So, for a constant
Weyl scaling
\be
\delta F=\int d^4x\,{\sqrt g}
\delta\rho {\cal A}=-{\delta\rho\over 2}\left(\tau{\partial F\over\partial\tau}+\tau'{\partial F\over\partial\tau'}\right)
\label{der}
\ee

At one-loop we only need the 
quadratic fluctuations in the action, so the fields are essentially free.
In \cite{us} we computed the Weyl anomaly for free scalar and 
spin-half particles for the metric (\ref{newmet}), not by performing
a functional integration but
by interpreting
$\Psi_{\tau,\tau'}[\tilde\varphi,\varphi]$ (after Wick rotation of 
$g_{ij}$) as the \Sc functional, i.e. the
matrix element of
the time evolution operator between eigenstates of the field,
\be
\Psi_{\tau,\tau'}[\tilde\varphi,\varphi]
=\langle \,\tilde\varphi\,|
\, T\, \exp(-\int_{\tau'}^{\tau} d t\, H ( t) )\,|\,\varphi\,\rangle\, ,
\label{fun}
\ee
To illustrate this consider a massless scalar field. $\Psi$ satisfies the
functional Schr\"odinger equation
\be
{\partial\over \partial\tau}\Psi_{\tau,\tau'}[\tilde\varphi,\varphi]
={\textstyle 1\over 2}\int d\x\,\left (\tau^3\,{\delta^2\over\delta\tilde\varphi^2} +\Omega\, \tilde\varphi \nabla
\cdot\nabla\tilde\varphi +2\,\delta^4(0)/\tau\right)\Psi_{\tau,\tau'}[\tilde\varphi,\varphi],\label{Schr}
\ee
with a similar equation for the $\tau'$ dependence, and the initial condition that $\Psi$ becomes a delta-functional as $\tau'$ approaches $\tau$. The
logarithm of
$\Psi$ takes the form
\be
F+\int d^d\x\,\left({1\over 2}\tilde\varphi\,\Gamma_{\tau,\tau'}\,\tilde\varphi+\tilde\varphi\,\Xi_{\tau,\tau'}\,
\varphi+{1\over 2}\varphi\,\Upsilon_{\tau,\tau'}\,\varphi\right),
\label{expand'}
\ee
so that the \Sc equations relate the derivatives of $F$ with respect to
$\tau$ and $\tau'$ that appear in (\ref{der}) to the functional traces of
$\Gamma_{\tau,\tau'}$ and $\Upsilon_{\tau,\tau'}$. These operators
are readily obtained by solving the \Sc equations in powers of 
$\nabla
\cdot\nabla$ so that eventually we obtain $\cal A$
in terms of the heat-kernel associated with this four-dimensional operator.
We found that when a mass is included $\cal A$ vanishes for generic values of the mass. However, it is not a continuous function, and when the 
mass is such that $\sqrt{l^2m^2+4}$ is an integer, $N$, then
${\cal A}=-N\,R^{ijkl}R_{ijkl}/(5760\pi^2)$. Similarly, for spin-half
fermions, $\cal A$ is non-vanishing when $|lm|+1/2$ is an integer. 
These are precisely the mass values that occur in the spectrum of
Supergravity. 

\medskip

Although our original calculation only applied to scalar and spin-half
fields, it gives us most of the information we need to calculate $\cal A$
for
fields of {\em arbitrary} spin. The strategy is to decompose the fields in such
a way that they satisfy functional equations of a similar form as those
satisfied by 
their lower spin counterparts. For example, consider a massive vector field
$A_\mu$. 
The Lagrangian density, as usual, is proportional to the square of the field
strength. Thus $A_0$ is non-dynamical, and may be integrated out of the
partition function. Now
the {\em transverse} part of $A_i$ has a Lagrangian which is identical in
form to the scalar field Lagrangian, and the calculation we outlined above
holds for this field with only two differences: the heat-kernel of
$\del\cdot\del$ is the one acting on transverse vector fields, and the
inner-product on $A_i$ has a measure which differs by a factor of $t^2$ from
the inner-product on scalars; this has the effect that the mass values for
which the anomaly is non-zero are changed to $\sqrt{l^2m^2+1}=N$.

So what about the remaining longitudinal mode? The trick is to make a change
of variables so that the Jacobian cancels the unwanted determinant coming from
integrating out $A_0$. Now the resulting Lagrangian is not quite the usual
Lagrangian for scalar fields, but it is the same to lowest order in
$\del\cdot\del$, which is all we need to calculate the anomaly. The mass term
is such that the anomaly is again non-zero for $\sqrt{l^2m^2+1}=N$. 

Finally, the heat-kernels of $\del\cdot\del$ acting on a transverse vector
and a scalar sum to the heat-kernel of $\del\cdot\del$ acting on
an {\em unconstrained} vector. So the anomaly arising from a vector field in {\em
  five} dimensional AdS spacetime is obtained from the heat kernel of the
Laplacian acting on {\em four} dimensional vector fields. Furthermore, the
anomaly is non-zero for integer values of $\sqrt{l^2m^2+1}$, which are again
precisely the mass values appearing in the Supergravity spectrum. 

The same thing happens for all the fields of the theory. For generic values
of the masses 
the Weyl anomalies vanish, but for those values that occur in the
spectrum $\cal A$ is non-zero. Now there is a small ambiguity in the calculation of the anomaly related to the choice of boundary conditions. (For bosonic fields we can choose a combination of Dirichlet and Neumann conditions, and for fermionic fields we 
have a choice of projection for the boundary operators). In each case there is a special choice which leads to a mass independent correction relative to the generic choice. 

We find that to make the correpondence work we need to make the generic choice. For the fermions this leads to a shift $N\-N-1/2$ relative to the calculation described in \cite{us}. With this stipulation, the anomaly is given for bosonic
fields by
\be {\cal A}=-\alpha\,N\,R^{ijkl}R_{ijkl}/(5760\pi^2), \ee
where $\alpha$ is the coefficient coming from the relevant heat kernel,
normalized to 1 for scalar fields. For fermionic fields we
have
\be {\cal A}=-\alpha\,(N-1/2)\,R^{ijkl}R_{ijkl}/(5760\pi^2). \ee
We
list the values of $N$ and $\alpha$ for all the massive fields in table \ref{coeffs}.
Although our test of the Maldacena
conjecture could have been passed trivially by all the one-loop
contributions vanishing separately, we have instead non-vanishing contributions
from each of the infinite number of fields corresponding to the
Kaluza-Klein modes on $S^5$, and we will observe a highly non-trivial cancellation between them.

\medskip

Now the mass spectrum is given in table \ref{spec}, where the supermultiplets
are labelled by an integer $p$. The first three fields in the table form the
doubleton representation for $p=1$, but this is not present in the
spectrum. The $p=2$ or massless multiplet contains all the gauge fields,
which we will discuss in a moment. For the $p\ge3$ multiplets the total
anomaly is proportional to $\sum_{\rm fields}\alpha Nr$ where $r$ is the
dimension of the $SU(4)\sim SO(6)$ representation in which the fields
transform. It can be seen from the table that $r$ is always a polynomial of degree 4
in $p$, while $N$ is linear in $p$. So each field contributes a polynomial of
degree 5 to the sum. If we add up the contributions from all the fields, we
find that the anomaly vanishes for each of the $p\ge3$ multiplets.

Before we discuss the massless multiplet, notice that the result so far is
in
accordance with the existence of a consistent truncation which essentially
discards all of the $p\ge3$ multiplets. Even if the overall anomaly was non-zero,
the contribution from these multiplets should still vanish.

So we turn to the $SO(6)$ gauge fields of the massless multiplet. The story
is still much the same as before, but we need to fix the gauge symmetries and
include Faddeev-Popov ghosts. In table \ref{ghosts} we show how the
decomposition works for each of the gauge fields. Using this information it
is easy to see that the anomaly sums to zero for this multiplet also, so that
the Maldacena conjecture has successfully passed our test.

\begin{table}[tbp]
\begin{center}
\caption{Anomaly coefficients of massive fields on $AdS_5$.}
\label{coeffs}
\vskip .3cm
\begin{tabular}{|ccc|}
\hline
Field & $N$ & $\alpha$\\
\hline
$\phi$ & $\sqrt{m^2+4}$ & 1\\
$\psi$  & $|m|+1/2$ & 7/2\\
$A_\mu$ & $\sqrt{m^2+1}$ & -11\\
$A_{\mu\nu}$ & $m$ & 33\\
$\psi_\mu$ & $|m|+1/2$ & -219/2\\
$h_{\mu\nu}$ & $\sqrt{m^2+4}$ & 189\\
\hline

  \end{tabular}
 \end{center}
\end{table}

\begin{table}[tbp]
\begin{center}
\caption{Mass spectrum. The supermultiplets (irreps of U(2,2/4)) are labelled
  by the integer $p$. Note that the doubleton ($p=1$) does not appear in the
  spectrum. The $(a,b,c)$ representation of $SU(4)$ has dimension
  $r=(a+1)(b+1)(c+1)(a+b+2)(b+c+2)(a+b+c+3)/12$, and a subscript $c$ indicates
  that the representation is complex. (Spinors are four component Dirac
  spinors in $AdS_5$).}                            
\label{spec}
\vskip .3cm
 \begin{tabular}{|cccc|}
\hline

    Field  & $SO(4)$ rep$^{\rm n}$ & $SU(4)$ rep$^{\rm n}$ & Mass on $S^5$      \\

 \hline

$\phi^{(1)}$ & $(0,0)$ & $(0,p,0)$ & $m^2=p(p-4)$,\quad $p\ge2$ \\
$\psi^{(1)}$ & $(\hf,0)$ & $(0,p-1,1)_c$ & $m=p-3/2$,\quad $p\ge2$ \\
$A_{\mu\nu}^{(1)}$ & $(1,0)$ & $(0,p-1,0)_c$ & $m^2=(p-1)^2$,\quad $p\ge2$ \\
\hline
$\phi^{(2)}$ & $(0,0)$ & $(0,p-2,2)_c$ & $m^2=(p+1)(p-3)$,\quad $p\ge2$ \\
$\phi^{(3)}$ & $(0,0)$ & $(0,p-2,0)_c$ & $m^2=(p+2)(p-2)$,\quad $p\ge2$ \\
$\psi^{(2)}$ & $(0,0)$ & $(0,p-2,1)_c$ & $m=p-1/2$,\quad $p\ge2$ \\
$A_\mu^{(1)}$ & $(\hf,\hf)$ & $(1,p-2,1)$ & $m^2=p(p-2)$,\quad $p\ge2$ \\
$\psi_\mu^{(1)}$ & $(1,\hf)$ & $(1,p-2,0)_c$ & $m=p-1/2$,\quad $p\ge2$ \\
$h_{\mu\nu}$ & $(1,1)$ & $(0,p-2,0)$ & $m^2=(p+2)(p-2)$,\quad $p\ge2$ \\
\hline
$\psi^{(3)}$ & $(\hf,0)$ & $(2,p-3,1)_c$ & $m=p-1/2$,\quad $p\ge3$ \\
$\psi^{(4)}$ & $(\hf,0)$ & $(0,p-3,1)_c$ & $m=p+1/2$,\quad $p\ge3$ \\
$A_\mu^{(2)}$ & $(\hf,\hf)$ & $(1,p-3,1)_c$ & $m^2=(p+1)(p-1)$,\quad $p\ge3$ \\
$A_{\mu\nu}^{(2)}$ & $(1,0)$ & $(2,p-3,0)_c$ & $m^2=p^2$,\quad $p\ge3$ \\
$A_{\mu\nu}^{(3)}$ & $(1,0)$ & $(0,p-3,0)_c$ & $m^2=(p+1)^2$,\quad $p\ge3$
\\
$\psi_\mu^{(2)}$ & $(1,\hf)$ & $(1,p-3,0)_c$ & $m=p+1/2$,\quad $p\ge3$ \\
\hline 
$\phi^{(4)}$ & $(0,0)$ & $(2,p-4,2)$ & $m^2=(p+2)(p-2)$,\quad $p\ge4$ \\
$\phi^{(5)}$ & $(0,0)$ & $(0,p-4,2)_c$ & $m^2=(p+3)(p-1)$,\quad $p\ge4$ \\
$\phi^{(6)}$ & $(0,0)$ & $(0,p-4,2)$ & $m^2=p(p+4)$,\quad $p\ge4$ \\
$\psi^{(5)}$ & $(\hf,0)$ & $(2,p-4,1)_c$ & $m=p+1/2$,\quad $p\ge4$ \\
$\psi^{(6)}$ & $(\hf,0)$ & $(0,p-4,1)_c$ & $m=p+3/2$,\quad $p\ge4$ \\
$A_\mu^{(3)}$ & $(\hf,\hf)$ & $(1,p-4,1)$ & $m^2=p(p+2)$,\quad $p\ge4$ \\
\hline 
 \end{tabular}
 \end{center}
\end{table}

\begin{table}[tbp]
\begin{center}
\caption{Decomposition of gauge fields for the massless multiplet. Some components of the graviton have
  non-physical mass values, but cancel against identical Faddeev-Popov
  ghosts. A superscript ``irr'' indicates that the corresponding field is
  irreducible.}
\label{ghosts}
\vskip .3cm
\begin{tabular}{|c|ccc|}
\hline
Original field & Gauge fixed fields & $N$ & $\alpha$\\
\hline
$A_\mu$ & $A_i$ & 1 & -11\\
({\bf 15} of $SU(4)$)       & $A_0$ & 2 & 1\\
        & $b_{FP}$, $c_{FP}$ & 2 & -1\\
\hline
$\psi_\mu$ & $\psi_i^{\rm irr}$ & 2 & -219/2\\
&            $\gamma^i\psi_i$ & 3 & 7/2\\
({\bf 4} of $SU(4)$)  & $\psi_0$ & 3 & 7/2\\
& $\lambda_{FP}$, $\rho_{FP}$ & 3 & -7/2\\
& $\sigma_{GF}$ & 3 & -7/2\\
\hline
$h_{\mu\nu}$ & $h_{ij}^{\rm irr}$ & 2 & 189\\
($SU(4)$ singlet) & $h_{0i}$ & 3 & -11\\
& $h_{00}$, $h_\mu^\mu$ &cancel& $B^{FP}_0$,$C^{FP}_0$\\
& $B^{FP}_0$,$C^{FP}_0$ &cancel& $h_{00}$, $h_\mu^\mu$\\
& $B^{FP}_i$,$C^{FP}_i$ & 3 & 11\\
\hline 
  \end{tabular}
 \end{center}
\end{table}

To summarize. The coefficient of the square of the Riemann tensor in the 
Weyl anomaly of ${\cal N}=4$ Super Yang-Mills theory is zero to all loops.
For the Maldacena conjecture to hold the same must be true in the
calculation on the $AdS_5$ side. We found that at one-loop the 
contributions
of the individual species of fields was non-zero, however they sum
to zero over each supermultiplet, in agreement with the conjecture.
There is a similar cancellation over supermultiplets
in the case of the chiral anomaly \cite{chu}. To make the cancellation work required that we make a generic choice of boundary conditions for the bulk fields; if we had chosen certain special boundary conditions, the calculation would have failed, leading
 to pathological divergences in the boundary theory.

Our calculation was done in the field theory limit of String
Theory, however we expect that the neglected string modes
have contributions  that vanish separately. This is because
the anomaly is not a continuous function of mass, but vanishes for
generic values. It is non-zero for the Kaluza-Klein spectrum of supergravity
which has a special property. A standard construction of Anti-de-Sitter space
is as a hyperboloid in $R^{2,4}$. This contains closed time-like
curves which are removed by going to the universal cover.
Curiously the spectrum of supergravity allows the corresponding
fields to be single-valued on this hyperboloid. This property
will not be shared by the higher string modes whose masses depend on the
string-scale $\alpha'$. 

We have calculated one half of the Weyl anomaly. The
remaining terms that depend on the Ricci tensor can be computed by a
similar calculation but in a different background, say one 
corresponding to a spherical boundary, for which the Einstein equations can
again be solved exactly. This would provide a further test,
as would the computation of higher loops in the bulk.

An essential part of our calculation was the use of the functional
\Sc equation which we consider to be a useful tool in the 
study of the AdS/CFT correspondence, not least because the central 
object of study in the correspondence is itself the Schr\"odinger functional.

\vfill\eject


\begin{thebibliography}{88}


\bibitem{Henningson} M. Henningson and K. Skenderis JHEP 9807 (1998), 023


\bibitem{Witten} E. Witten, Adv.Theor.Math.Phys.2 (1998), 253.


\bibitem{Maldacena} J. Maldacena, Adv.Theor.Math.Phys.2 (1998), 231.


\bibitem{Kim}
H.J.~Kim, L.J.~Romans and P.~van Nieuwenhuizen,
Phys. Rev. D32 (1985), 389. 

\bibitem{us}
P. Mansfield and D. Nolland,
JHEP 9907 (1999), 028.

\bibitem{chu}
A. Bilal and C-S. Chu,
Nucl. Phys. B562 (1999), 181.

\bibitem{blau}
M. Blau, K.S. Narain and E. Gava, JHEP 9909 (1999), 018.

\bibitem{odinsov}
S. Nojiri and S.D. Odintsov, Int. J. Mod. Phys. A15 (2000), 413.

\end{thebibliography}
\end{document}